# Quantum description of spherical spins

Th. M. Nieuwenhuizen
*Van der Waals-Zeeman Laboratorium, Universiteit van Amsterdam*
*Valckenierstraat 65, 1018 XE Amsterdam, The Netherlands*

The spherical model for spins describes ferromagnetic phase transitions well, but it fails at low temperatures. A quantum version of the spherical model is proposed. It does not induce qualitative changes near the phase transition. However, it produces a physical low temperature behavior. The entropy is non-negative. Model parameters can be adapted to the description of real quantum spins. Several applications are discussed. Zero-temperature quantum phase transitions are analyzed for a ferromagnet and a spin glass in a transversal field. Their crossover exponents are presented.

The spherical model was introduced by the late Marc Kac, see [1] for a discussion on its history and the route to its solution, presented by Berlin and Kac in 1952. [2] It was meant to be a simple soluble example for spin systems, from which the behavior near the phase transition can be studied in a simple manner. A review of this subject was given by Joyce. [3]

The model also has a drawback: the behavior in the low temperature phase is pathological. The specific heat is just constant, while the entropy goes to minus infinity as $S \sim \log T$. These aspects even occur for free spherical spins in a magnetic field. For these reasons the practical use of the spherical model has been limited.

We have wondered whether the low $T$ behavior can be healed by modifying the spherical model. We have considered the definition of spherical spins in a standard thermal field theory. We report our findings here.

We consider spins coupled ferromagnetically to nearest-neighbors in the presence of a field

$$\mathcal{H} = -J \sum_{<ij>} \mathbf{S}_i \mathbf{S}_j - \mathbf{H} \sum_i \mathbf{S}_i \tag{1}$$

The $\mathbf{S}_i$ are $m$-component variables. We shall mainly consider $m = 3$-component vector spins. The spins are subject to the spherical constraint

$$\frac{1}{N} \sum_{i=1}^{N} \mathbf{S}_i^2 = \sigma \tag{2}$$

Later we will specify $\sigma$. The classical partition sum reads

$$Z = \int D\mathbf{S}\, e^{-\beta\mathcal{H}} \tag{3}$$

where $D\mathbf{S} = \delta(\sum_{i=1}^{N} \mathbf{S}_i^2 - N\sigma) \prod_i \{d\mathbf{S}_i/(2\pi)^{m/2}\}$. For free spins ($J = 0$) in a field one finds the free energy

$$F = -\frac{\mu\sigma}{2} + \frac{mT}{2} \log \beta\mu - \frac{H^2}{2\mu} \tag{4}$$

The magnetization is $M = H/\mu$. Variation with respect to the "chemical potential" $\mu$ yields the equation of state $M^2 = \sigma - mT/\mu$. It can be solved explicitly

$$M = \frac{2\sigma H}{mT + \sqrt{m^2T^2 + 4\sigma H^2}} \tag{5}$$

The susceptibility, $\chi = \sigma/mT$, follows a perfect Curie law. In fixed field the magnetization goes to $\sqrt{\sigma}$ as $T \to 0$. However, the entropy goes to $-\infty$ as $(m/2)\log T$ for $T \to 0$. This is in conflict with the third law of thermodynamics, but occurs always for classical vector spins. The specific heat $C = TdS/dT$ goes to a constant.

We wish to improve this low temperature behavior. If we return to eq. (1) at $H = 0$ and denote the eigenvalues of the coupling matrix by $\lambda$, we obtain $Z \sim \max_\mu \exp(Nm\beta\mu/2) \prod_\lambda [\beta(\mu - \lambda)]^{-m/2}$. In analogy with harmonic oscillators, we expect that the term $\beta(\mu - \lambda)$ arises from a quantum expression $\sinh \alpha\beta(\mu - \lambda)$. Such a factor can indeed be derived when no square root occurs. We therefore propose two ways to quantize the spherical model. If spins are real valued, the quantization must be defined for $Z^2$. This is reminiscent of the solution of the 2-d Ising model, where $Z$ can be evaluated with Pfaffians, while $Z^2$ can be expressed in terms of Grassmann integrals. If the spins are complex, the quantization can be defined for $Z$ itself.

For real valued spins, one doubles its components. One thus considers spins $\mathbf{S}_i^a(\tau)$ ($a = 1,2$) defined on an "imaginary time" interval $0 < \tau < \beta$, subject to periodic boundary conditions $\mathbf{S}_i^a(\beta) = \mathbf{S}_i^a(0)$. In units where $\hbar = 1$ the square of the partition sum reads

$$Z^2 = \int \mathcal{D}\mathbf{S} \exp \int_0^\beta d\tau \left\{ \frac{-1}{4\alpha} \sum_{i=1}^{N} \sum_{a,b=1}^{2} \mathbf{S}_i^a(\tau) \sigma_{ab}^{(2)} \frac{d}{d\tau} \mathbf{S}_i^b(\tau) - \sum_{a=1}^{2} \mathcal{H}(\{\mathbf{S}_i^a(\tau)\}) \right\} \tag{6}$$

where the $\tau$-integral is discretized in $\mathcal{M} = \beta/d\tau$ steps and $dS(\tau) = S(\tau + d\tau) - S(\tau)$. The two spins systems are coupled by the Pauli-matrix in the derivative term, viz. $\sigma_{21}^{(2)} = -\sigma_{12}^{(2)} = i$; $\sigma_{11}^{(2)} = \sigma_{22}^{(2)} = 0$. The integration measure is a normalization constant times the repetition of the sperical measure at all imaginary time points, $\mathcal{D}\mathbf{S} = C^{Nm} \prod_{a,\tau} D\mathbf{S}^a(\tau)$ with $C = (1/2\alpha) \prod_{n \neq 0} (\pi|n|/2\alpha)$.

The quantum term in (6) involves a first order derivative in time. Our motivation hereto was that $d$-dimensional quantum spins systems, such as the Ising chain in a transverse field, involve such a term due to the Trotter-Suzuki transformation into a classical $d+1$-dimensional problem. In a recent work on the spherical model other systems were considered where a second order derivative occurs. [4] For magnetic systems this does not lead to acceptable low temperature behavior.

In order to fix some universal properties of the system, we consider again free spins in a field. At each $\tau$ we introduce a Lagrange multiplier $\mu(\tau)$ for the spherical constraint. The mean field value occurs for constant $\mu$. If the thermodynamic limit $N \to \infty$ is taken before the limit $\mathcal{M} \to \infty$, *the fluctuations do not yield extensive contributions.* We thus find simply, instead of eq. (4),

$$F = -\frac{\mu\sigma}{2} + \frac{mT}{2} \log 2 \sinh \alpha\beta\mu - \frac{H^2}{2\mu} \tag{7}$$

Variation with respect to $\mu$ now yields

$$M^2 = \frac{H^2}{\mu^2} = \sigma - m\alpha \coth \alpha\beta\mu \tag{8}$$

For large $T$, $\mu$ grows linearly, $\mu = (y/\alpha)T$ with $\coth y = \sigma/m\alpha$. The susceptibility follows the Curie-law $\chi = 1/\mu = \alpha/yT$. The entropy reads

$$S = \frac{m}{2}(\alpha\beta\mu \coth \alpha\beta\mu - \log 2 \sinh \alpha\beta\mu) \tag{9}$$

Whereas in the classical model there was no unique way to define its adjustable constant, we have chosen it here such that, in a finite field, $S$ vanishes at $T = 0$. In the classical limit $\alpha \sim \hbar^2 \to 0$ one might be tempted to neglect the term $-(m/2)\log \alpha$. It is exactly this procedure, however, that produces a negative classical entropy.

In zero field, or at infinite $T$, the entropy becomes

$$S_\infty = \frac{m}{2} \left( \frac{y\sigma}{m\alpha} - \log 2 \sinh y \right) \tag{10}$$

The zero-point magnetization $M_0 = \sqrt{\sigma - m\alpha}$ exhibits a quantum reduction from the classical value $\sqrt{\sigma}$. Eqs. (7) and (9) exhibit an energy gap $\Delta E = 2\alpha\mu_0 = 2\alpha H/M_0$. This is indeed expected for free quantum spins.

We now wish to see in how far our spherical model can be used to describe realistic quantum spins. In our definition of the thermal field theory we had two parameters: the length $\sigma$ and $\alpha \sim \hbar^2 \equiv 1$. We can fix these parameters by comparing to two expressions from the high and/or low temperature thermodynamics of free spins. Let us therefore consider quantum spins of order $S$ where $S = 1/2, 1, 3/2, \cdots$ is half-integer or integer. Our field $H$ is the product of the external field $H_0$, the $g$-factor and the Bohr magneton $\mu_B$, viz. $H = g\mu_B H_0$. For free quantum spins the susceptibility follows a Curie law

$$\chi_0 = g^2 \mu_B^2 \chi = \frac{g^2 \mu_B^2 S(S+1)}{3T} \tag{11}$$

This law is reproduced by our model with $m = 3$ if $\tanh 3\alpha/S(S+1) = 3\alpha/\sigma$. We can find a second constraint by adjusting the zero point magnetization to $M_0 = S$. This implies $\sigma = S^2 + 3\alpha$. Combining these relations one has to solve

$$\tanh y = \frac{(S+1)y}{S+(S+1)y} \tag{12}$$

This equation has a solution for any positive $S$. We find the values, presented in the table. For large $S$, $\sigma$ is close to $S(S+1)+1/3$, while $\alpha \approx S/3+1/9$. The energy gap, $\Delta E/H \approx 2/3+2/9S$, is below the value $\Delta E/H = 1$ of spin-$S$ quantum spins, but it is bounded by 2/3.

In Figure 1 we present the specific heat of a free $S=1$ spin with the values of $\sigma$ and $\alpha$ taken from the Table. It is compared to the "classical" result (same $\sigma$; $\alpha \to 0$), which goes to 3/2 at $T=0$. We have also presented the specific heat of a $S=1$ quantum spin, viz. $C = (\beta H/2\sinh(\beta H/2))^2 - (3\beta H/2\sinh(3\beta H/2))^2$. This expression vanishes quicker at small $T$, since it has a larger gap than our result for the free spherical spin.

Having fixed the system parameters we can now consider coupled spins. The first case is the mean field ferromagnet, eq. (1) with the sum involving all pairs. We denote $J_0 = NJ$. The free energy reads

$$F = -\frac{\mu\sigma}{2} + \frac{mT}{2}\log 2\sinh\alpha\beta\mu + \frac{J_0 M^2}{2} - \frac{(J_0 M + H)^2}{2\mu} \tag{13}$$

When $M \neq 0$ in zero field, one has $\mu = J_0$ and

$$M^2 = \sigma - \alpha m \coth\alpha\beta J_0 \tag{14}$$

At the phase transition $M$ grows as $(T_c - T)^\beta$ with critical exponent $\beta = 1/2$, as in the classical description. The entropy in the ordered phase

$$S = \frac{m}{2}\left[\alpha\beta J_0\left(\coth\alpha\beta J_0 - 1\right) - \log\left(1 - e^{-2\alpha\beta J_0}\right)\right] \tag{15}$$

vanishes exponentially at low $T$. A similar behavior occurs for Ising spins with mean field couplings, since the latter just act as an external field, of adjusted strength.

It is known that the spherical spins behave differently in case of short range ferromagnetic interactions. In $d$-dimensions one finds

$$F = -\frac{\mu\sigma}{2} + \frac{mT}{2}\int\frac{d^dk}{(2\pi)^d}\log 2\sinh\alpha\beta[\mu - J(k)] - \frac{H^2}{2(\mu - J_0)} \tag{16}$$

where $J_0 = J(k=0)$. Our quantum formulation brings no qualitative changes near the phase transition. One still has $\beta = 1/2$, $\eta = 0$, while also the other exponents are unchanged. However, near the transition the low temperature branch of the specific heat is linear rather than constant. At low $T$ the excitations are spin waves. Their degeneracy is $m$, and not $m-1$, because, apart from the direction, also the length of the spins can fluctuate. The entropy and the specific heat decay as $mT^{d/2}$ at small $T$. In Figure 2 we present a plot of the specific heat for $S=1$ spherical "Heisenberg" spins on a simple cubic lattice, with $\sigma$ and $\alpha$ taken from the table. It is compared to the classical result ($\alpha = 0$).

Another case of interest is when only the $z$-components of the spins have an exchange coupling. The field may have a component in the transverse direction; a well known example is the Ising chain in a transverse magnetic field. For $m=3$-component spherical spins one finds the free energy

$$\begin{aligned} F = &-\frac{\mu\sigma}{2} + \frac{T}{2}\int\frac{d^dk}{(2\pi)^d}\log 2\sinh\alpha\beta[\mu - J(k)] \\ &+ T\log 2\sinh\alpha\beta\mu - \frac{H_\perp^2}{2\mu} - \frac{H_z^2}{2(\mu - J_0)} \end{aligned} \tag{17}$$

In a perpendicular field it leads to the equation of state

$$M_z^2 = \sigma - \alpha\int\frac{d^dk}{(2\pi)^d}\coth\alpha\beta[\mu - J(k)] - 2\alpha\coth\alpha\beta\mu - \frac{H_\perp^2}{\mu^2} \tag{18}$$

For $d > 2$ a phase transition occurs where $M_z \neq 0$. The ordering is suppressed when $H_\perp$ is large enough. At $T=0$ there occurs a quantum phase transition at critical field $H_\perp = J_0 S$. Generally, there is an energy gap, $2\alpha H_z/M_z$; it only vanishes in the ordered phase at zero field. There is a scaling region when $\Delta = J_0 S - H_\perp$, $H_z$, $T$, and $\beta H_z/M_z$ are small. Here the singular part of the free energy and the order parameter assume the scaling form $F_{sing} = \Delta^2\Phi_1(x,y)$, $M_z = \Delta^{1/2}\Phi_2(x,y)$ with $x = H_z/|\Delta|^{\phi_H}$, $y = T/|\Delta|^{\phi_T}$ with crossover exponents $\phi_H = 2/3$, $\phi_T = 2/d$. At non-zero $H_z$ and very small $T$ the thermal excitations experience the gap.

A further subject of interest is the limit of strong anisotropy for Heisenberg spins. For considering the uniaxial limit where the spins point in the $z$-direction (Ising-like) we add an anisotropy term

$$\mathcal{H}_{ani} = D\sum_i S_{ix}^2 + S_{iy}^2 \tag{19}$$

to the Hamiltonian. For large $D$ the free energy reduces to eq. (16) with $m = 1$, provided we subtract the 'zero point energy' $2\alpha D$ and make the shift $\sigma \to \tilde{\sigma} = \sigma - 2\alpha$. Under these modifications we can fully neglect the orthogonal degrees of freedom. The zero point magnetization remains equal to $M_0 = S$. When first $D \to \infty$, the $T = \infty$ entropy reduces to

$$S_\infty = \frac{y}{2\alpha}\tilde{\sigma} - \frac{1}{2}\log 2\sinh y \tag{20}$$

This is below the isotropic value, as it should.

In systems with XY-spins, the $z$-component of the spin is suppressed. Adding a term $D\sum_i(S_{iz}^2 - \alpha)$ we find back eq. (16) with $m = 2$, provided we take $\sigma \to \tilde{\sigma} = \sigma - \alpha$.

Next we investigate quantum effects in the mean field spherical spin glass. This system is described by eq. (1) where now for each pair the coupling $J_{ij}$ is an independent Gaussian random variable with average zero and variance $J_2^2/N$. The model was introduced by Kosterlitz, Thouless and Jones [6], and extended to the situation with short range ferromagnetic interactions by the present author [7]. In zero field the quantum free energy is given by eq. (16), provided we replace $\int d^dk/(2\pi)^d$ by $\int dJ\sqrt{4J_2^2 - J^2}/2\pi J_2^2$, involving the semi-circular eigenvalue density of the random coupling matrix. There is a phase transition where $\mu$ sticks at $2J_2$. Here the Edwards-Anderson parameter $q$ becomes non-zero. At low $T$ the entropy and specific heat now vanish as $T^{3/2}$. In this mean field model there is no gap.

For a spherical spin glass with couplings only in the $z$-direction, we can again look at the effect of a perpendicular field, while $H_z = 0$. We find a similar analog of eq. (17), while in eq. (18) $q$ should replace $M_z^2$. At $T = 0$ there is a quantum phase transition for $H_\perp = J_0 S$, where the spin glass order is suppresed. In the disordered phase $\Delta < 0$, there is again a gap. In the regions where the gap vanishes or can be neglected, the singular part of the free energy and the order parameter have the scaling form $F_{sing} = \Delta^2\Phi_1(T/|\Delta|^{2/3})$, $q = \Delta\Phi_2(T/|\Delta|^{2/3})$.

For complex valued spherical spins we consider the thermal partition sum

$$Z = \int \mathcal{D}\mathbf{S}\exp\int_0^\beta d\tau\left\{\frac{-1}{4\alpha}\sum_{i=1}^N \mathbf{S}_i^*(\tau)\frac{d}{d\tau}\mathbf{S}_i(\tau) - \mathcal{H}(\mathbf{S}_i(\tau))\right\} \tag{21}$$

with $\mathcal{D}S = C^N\prod_\tau DS(\tau)$ and $DS = \delta(2N\sigma - \sum_i S_i^* S_i)\prod_i(dS_i^* dS_i/2\pi)$ being the spherical measure for complex spins. The Hamiltonian now reads

$$\mathcal{H} = -\frac{1}{2}J\sum_{i,j} S_i^* S_j - H\sum_i(\mathrm{Re}S_i + \mathrm{Im}S_i) \tag{22}$$

We find the same results as above, provided we identify the magnetization $M$ with $(-1/2)\partial F/\partial H$. Other couplings to the external field or interpretations of the magnetization lead to quantitative differences.

So far all models discussed have Gaussian properties since spins either are coupled in pairs or to an external field. However, in ref. [8] we have also allowed for additional random couplings between quartets of spins and possibly higher spin multiplets. It was found that in this situation replica symmetry is broken. Near the phase transition there appears a very close analogy with the SK model [9] for the mean field Ising spin glass. The replica symmetry breaking solution could be evaluated explicitly. The order parameter function has a unique shape at all $T$. In the classical description this model has again the unphysical low temperature effects of the spherical model. As, to our knowledge, it is the first spin glass model with infinite order replica symmetry breaking for which the whole low temperature phase has been solved exactly, it is an important question whether quantization using eq. (21) then leads a physically acceptable behavior in at low temperatures. It could indeed be shown that the entropy vanishes at $T = 0$. In fact, both entropy and specific heat again vanish as $T^{3/2}$ at low $T$.

In conclusion, we have discussed a quantum description of the spherical model. For real spins we quantize $Z^2$, for complex spins $Z$ itself. The low temperature behavior is physical. Free spins in a field behave at low $T$ indeed as quantized objects. The same occurs in case of mean field couplings, since they lead to an effective field. For short range couplings the excitations are spin waves. In systems with uniaxial couplings the order will be suppressed in a large enough transversal field. The behavior near zero temperature quantum phase transitions is analyzed. Also uniaxial (Ising-like) and easy plane (XY-spins) systems can be described in appropriate limits.

There occur two system parameters that can be adjusted to describe the correct Curie law at large temperatures and the correct zero point magnetization of real quantum spins. It is hoped that the spherical model can now be used in practical situations.

## ACKNOWLEDGMENTS

This work is dedicated to the memory of Marc Kac. It is a pleasure to thank Bernard Nienhuis and Mark van Rossum for discussion. This work was made possible by the Royal Dutch Academy of Arts and Sciences (KNAW).

---

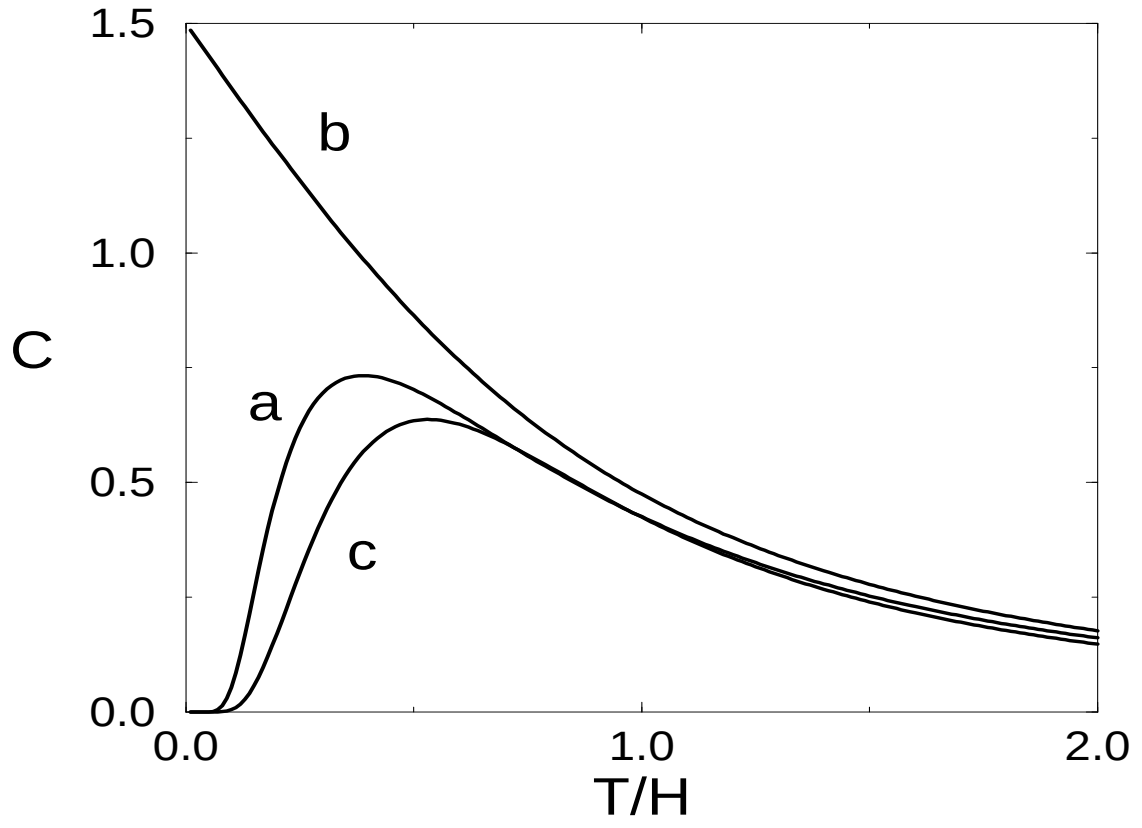

FIG. 1. Specific heat of a free spin. a) quantum spherical model; b) classical spherical model; c) quantum $S = 1$ spin.

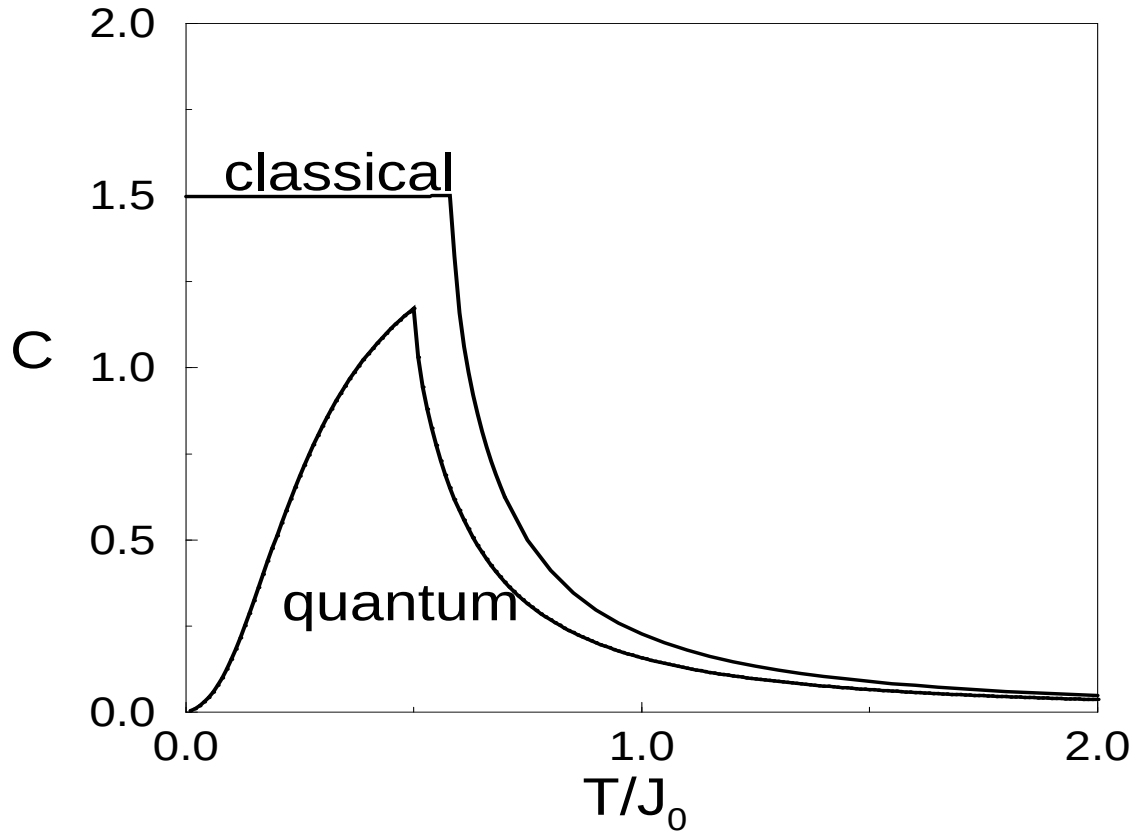

FIG. 2. Specific heat of $S = 1$ spherical spins on a simple cubic lattice: classical versus quantum situation.

| $S$ | $M_0$ | $\alpha$ | $\sigma$ | $\Delta E/H$ |
|---|---|---|---|---|
| 1/2 | 1/2 | 0.23797 | 0.96392 | 0.95190 |
| 1 | 1 | 0.41881 | 2.25643 | 0.83762 |
| 3/2 | 3/2 | 0.59213 | 4.02639 | 0.78950 |
| 2 | 2 | 0.76269 | 6.28807 | 0.76269 |
| 5/2 | 5/2 | 0.93192 | 9.04576 | 0.74553 |
| 3 | 3 | 1.10041 | 12.30121 | 0.73360 |
| 7/2 | 7/2 | 1.26842 | 16.05528 | 0.72481 |
| 4 | 4 | 1.43614 | 20.30843 | 0.71807 |
| 9/2 | 9/2 | 1.60365 | 25.06096 | 0.71273 |
| 5 | 5 | 1.77100 | 30.31300 | 0.70840 |

Table caption: Parameters of the spherical model.